\documentstyle[preprint,aps,epsfig]{revtex}
\widetext
\draft
\tighten

\begin{document}
%\pagenumbering{ }

\title{Experimental test of the compatibility  of the definitions of the
electromagnetic energy density and the Poynting vector}

\bigskip

\author{{\bf Andrew  Chubykalo, Augusto Espinoza and Rumen Tzonchev}} \address{{\rm Escuela de F\'{\i}sica, Universidad
Aut\'onoma de Zacatecas \\ Apartado Postal C-580,\\ Zacatecas 98068, ZAC.,
M\'exico\\ e-mails: andrew\_chubykalo@terra.com.mx,
agarrido@cantera.reduaz.mx and rumen@ahobon.reduaz.mx}}

%\date{\today}

\maketitle

\baselineskip 7mm

\bigskip

%\begin{center}
%Received $\;\;\;\;\;\;\;\;\;\;\;\;\;\;\;\;\;\;$ 2003
%\end{center}

\begin{abstract}
It is shown that the generally accepted definition of the Poynting vector
and the energy flux vector defined by means of  the energy density of the
electromagnetic field (Umov vector) lead to  the prediction of the 
different results touching  electromagnetic energy flux.  The experiment 
shows that within the framework of the mentioned  generally accepted 
definitions the Poynting vector adequately describes the electromagnetic 
energy flux unlike the Umov vector. Therefore one can conclude that a 
generally accepted definitions of the electromagnetic energy density and  
the Poynting vector, in general, are not always compatible.  
\end{abstract}

\bigskip

keywords: Poynting vector, energy density


\clearpage

\section{Introduction}

In the article ``Motion equations of the energy in the
bodies" [1] that appeared in the year that ``Tractate" was published
by Maxwell (1873), Umov developed the consequences from the
idea of the energy localization in the mediums. To each volume element in
the medium, the particles of which are in movement,  an
energy, constituted by the sum of the alive forces of the particles and
elements and the potential energy,   is associated. Umov thinks about the
problem of settling down in general form ``the laws of the transition of
the energy from an element to another", and to determinate, starting from
general principles, the study of the movement of the energy in the
mediums.  Starting from the energy conservation law Umov deduces the
motion equation for the energy in the mediums. If we represent the energy
density in a given point of the medium by means of $w$,  and through
$v_{x}$, $v_{y}$ and $v_{z}$ the energy velocity components in this point,
then the energy density loss in that point in unit of time is determined
by the relationship

\begin{equation}
-\frac{\partial w}{\partial t}=\frac{\partial \left( wv_{x}\right) }{
\partial x}+\frac{\partial \left( wv_{y}\right) }{\partial y}+\frac{\partial
\left( wv_{z}\right) }{\partial z}.
\end{equation}

\begin{quotation}
``The expression (1), similar to the expression of the mass
conservation law in hydrodynamics, is the expression of the elementary
energy conservation law in the mediums", Umov writes. From
this expression it can be established ``the relationship among
the quantity of energy, that in unit time leaves toward the medium
through its frontier, and the change of the quantity of energy in the
medium".
\end{quotation}

This relationship is expressed with the integral expression (Umov theorem)

\begin{equation}
\int\!\!\!\int\!\!\!\int  \frac{\partial w}{\partial
t}dxdydz+\int\!\!\!\int wv_{n}d\sigma =0.
\end{equation}

The vector $w{\bf v}$ defines the energy flow which crosses, in the unit
time, the perpendicular to this vector unitary surface. This is the so-called
Umov vector.

The case of the electromagnetic field, as particular case of the Umov
theorem, and therefore of the Umov vector,  was studied by Poynting.

In the year 1884 J. Poynting  published the article [2] that contained
the previously mentioned Umov-Poynting theorem. In this work Poynting
independently arrives to the same point of view developed 10 years before
by Umov. Poynting writes:  \begin{quotation} ``If we recognize  the
continuity of the energy movement, that is to say we recognize that when
the energy disappears in some point and appears in other, it should pass
through the intermediate space, then we are obliged to reach the
conclusion that the surrounding medium contains at the least a portion of
the energy and that it is capable to transmit the energy from one point to
another." \end{quotation}

Further on Poynting, leaning on  the Maxwell idea about the energy
localization in the field, formulates in this way the main idea of his
work:

\begin{quotation}
``The objective of this article is to demonstrate that there
exists a general law for the energy transport, in agreement with which the
energy in any point moves perpendicularly to the plane containing the lines
of the electric and magnetic forces, and that the quantity of the energy
passing through the unitary surface in this plane, for unit of time, is
equal to the product of the magnitudes of these two forces multiplied by the
sinus of the angle among them and divided among $4\pi $".
\end{quotation}
By this way Poynting defines the energy-flux vector for the case of the
electromagnetic field.

Discussing today the  conception of the Poynting vector and the number of
 basic difficulties associated with this concept one can sense clearly
that neither among researchers (see, e.g., [3,4]  and corresponding
references there) nor among authors of the generally accepted text-books
of classical electrodynamics (see, e.g.  [5-9]) a general agreement
exists about the essence of the energy-flux vector related with
electromagnetic fields.  Actually, the well-known authors Panofsky and
Phillips state [5]:  \begin{quotation} {\small ``Paradoxical results may
be obtained if {\it one tries to identify} the Poynting vector with the
energy flow per unit area at any particular point"}.  \end{quotation}

Contrarily, Feynman states
[9] that {\it exclusively} the identification of the Poynting vector (in
its generally accepted form) with the energy flow per unit area allows
to understand the law of conservation of the angular momentum in some
special cases. Other well-known authors Landau and Lifshitz state [6]:
\begin{quotation}
{\small ``Therefore the integral $\oint
{\bf S}d{\bf f}$ must be interpreted as the flux of field energy across
the surface bounding the given volume, {\it so that the Poynting vector
${\bf S}$ is this flux density} -- the amount of field energy passing
through unit area of the surface in unit time."}
\end{quotation}
Tamm [7] also identifies  the Poynting vector with the energy flow per
unit area at any particular point, however, taking into account that the
definition ${\bf S}=\frac{c}{4\pi}({\bf E}\times{\bf H})$ is not unique. In
turn Jackson claims in his famous text-book [8]:  \begin{quotation}
``{\small The  vector ${\bf S}$, {\it representing energy flow}, is called
the Poynting vector. It is given by ${\bf S}={\bf E}\times{\bf H}$
(6.109) \footnote{Jackson uses SI}....  Relativistic considerations
(Section 12.10) show that (6.109) is {\it unique}."} \end{quotation}

The only way  we can  verify the standard formula for the
energy flow due to the electromagnetic field is by experiment. Feynman said
[9]:
\begin{quotation}
{\small ``There are, in fact, an infinite number  of possibilities for $w$
(energy density) and ${\bf S}$, and so far no one has thought of an
experimental way to tell which one is right."} \end{quotation}

In this work we theoretically rationalize that the Poynting vector (in its
standard definition) does not {\it always} coincide with the energy flux
vector (Umov vector) related with electromagnetic waves.
The results of the experiment show that the Poynting vector is not always
compatible with the generally accepted definition of the electromagnetic
energy density.

\section{Theoretical motivation of the experiment}

More often than not physicists implicitly suppose that the Poynting vector

\begin{equation}
{\bf S}=\frac{c}{4\pi}{\bf E}\times{\bf B}
\end{equation}
and Umov (energy flux) vector\footnote{The expression of the Umov vector $({\bf U}=w{\bf
v})$ is obtained from the general energy conservation law ($\frac{\partial w}{\partial t} = -\nabla\{w {\bf v}\}$) and describes the
energy flux density of {\it any} kind of energy (not only
electromagnetic energy), $w$ is the corresponding energy density  and
{\bf v} is the propagation velocity of the energy in a given point.  Thus
the Poynting and Umov vectors should always coincide.}

\begin{equation} {\bf U}=wv{\bf n}
\end{equation}
always coincide for {\it any} electromagnetic wave spreading in vacuum in
every point.  Here {\bf n} is a unit vector along the direction of
propagation of the electromagnetic energy, $v$ is  the transferring energy
velocity (in the case of electromagnetic waves in vacuum $v=c$)   and $w$
is the energy density of the electromagnetic wave.  In actual fact, this
assertion is shown at least for plane and spherical electromagnetic waves
in vacuum (see, e.g., [6], Eq. 47.5).  Nevertheless, the assertion that ${\bf
S}={\bf U}$ for waves of a more general kind  is not proved in textbooks
and monographs.

Let us study  what condition in vacuum for ${\bf E}$ and ${\bf B}$
in an electromagnetic wave must be satisfied when the equality ${\bf
S}={\bf U}$ is valid.   We have in CGS (Gauss' system):

\begin{equation}
{\bf S}=\frac{c}{4\pi}{\bf E}\times{\bf B}= \frac{c}{4\pi}EB
\sin\alpha{\bf n}
\end{equation}
and
\begin{equation}
{\bf U}=wc{\bf n}=\frac{c}{8\pi} (E^2+B^2){\bf n}.
\end{equation}
Equating (5) and (6) we obtain
\begin{equation}
2EB\sin\alpha=E^2+B^2
\end{equation}
or

\begin{equation}
(E-B)^2+2EB(1-\sin\alpha)=0.
\end{equation}
According to the problem definition we choose real values of $E$, $B$ and $\alpha$ only, where $\alpha$ is the angle between ${\bf E}$ and ${\bf B}$. Therefore the last
equality (8) can be valid if and only if $E=B$ and $\alpha=\pi/2$. Thus the 
{\bf Theorem} takes place: {\it for the equality of the Poynting vector and
Umov vector it is necessary and sufficient that ${\bf E}\bot{\bf B}$
and $E=B$}.

In the next sections we  propose and perform the experiment which
allows us to check the incompatibility of the conventional functional
forms of the  Poynting vector and the electromagnetic energy density when
the electromagnetic wave field does not satisfy the conditions
\begin{equation} {\bf E}\bot{\bf B}\qquad {\rm and}\qquad E=B.
\end{equation}

\section{Theoretical predictions}

In order to obtain theoretically the electromagnetic energy flow observed in the  experiment
described in  section IV, we model the sources with the help of two point
sources emitting spherical waves and we calculate the flow intensity by
using the definition of the electromagnetic energy flow according to the
Poynting vector and according to the Umov one.

Let the source 1 be placed in $\left( 0,-l,0\right) $ and the source 2 in
$\left( 0,l,0\right) $. The screen is placed in the plane $z=h$ with
$-a\leq x\leq a$ and $-b\leq y\leq b$ (Fig. 1.)

\begin{figure}
\caption{Position of the point sources and the screen.}
\end{figure}

The monochromatic spherical waves created by these sources can be modeled
with the following expressions for the electric  and magnetic fields:
\begin{equation}
{\bf E}_{1} =\frac{E_{0}}{R_{1}}\cos \left( {\bf k}_{1}\cdot
{\bf R}_{1}-\omega t\right) {\bf e}_{\theta 1},
\end{equation}

\begin{equation}
{\bf B}_{1} =\frac{B_{0}}{R_{1}}\cos \left({\bf k}_{1}\cdot
{\bf R}_{1}-\omega t\right) {\bf e}_{\varphi 1}
\end{equation}

\begin{equation}
{\bf E}_{2} =\frac{E_{0}}{R_{2}}\cos \left({\bf k}_{2}\cdot
{\bf R}_{2}-\omega t\right) {\bf e}_{\theta 2},
\end{equation}

\begin{equation}
{\bf B}_{2} =\frac{B_{0}}{R_{2}}\cos \left({\bf k}_{2}\cdot
{\bf R}_{2}-\omega t\right) {\bf e}_{\varphi 2},
\end{equation}
where ${\bf k}_{1}=\frac{k}{R_{1}}{\bf R}_{1}$ y ${\bf k}_{2}=
\frac{k}{R_{2}}{\bf R}_{2}$, $E_0$ and $B_0$ are amplitudes,
$k=2\pi/\lambda$ is wave number for both waves. By this way the energy
flow has radial direction for each of these sources \begin{equation} {\bf
S}_{1}=\frac{cE_{0}B_{0}}{4\pi R_{1}^{2}}\cos ^{2}\left({\bf k}_{1}\cdot
{\bf R}_{1}-\omega t\right) {\bf e}_{R1}, \end{equation}

\begin{equation}
{\bf S}_{2}=\frac{cE_{0}B_{0}}{4\pi R_{2}^{2}}\cos ^{2}\left({\bf
k}_{2}\cdot {\bf R}_{2}-\omega t\right) {\bf e}_{R2},
\end{equation}
where ${\bf e}_{R1}$, ${\bf e}_{\theta 1}$, ${\bf e}_{\varphi 1}$
and ${\bf e}_{R2}$, ${\bf e}_{\theta 2}$, ${\bf e}_{\varphi 2}$
are the corresponding local spherical orts associated to the sources 1 and
2. By means of $r$, $\theta $ y $\varphi $ we will designate the spherical
coordinates in our coordinate system. The corresponding electromagnetic
energy densities are defined by the expressions

\begin{equation}
w_{1}=\frac{E_{0}^{2}+B_{0}^{2}}{8\pi R_{1}^{2}}\cos ^{2}\left({\bf k}
_{1}\cdot {\bf R}_{1}-\omega t\right),
\end{equation}

\begin{equation}
w_{2}=\frac{E_{0}^{2}+B_{0}^{2}}{8\pi R_{2}^{2}}\cos ^{2}\left ({\bf
k}_{2}\cdot {\bf R}_{2}-\omega t\right) .
\end{equation}

If we take into account that $E_{0}=B_{0}$, then for each of the two
spherical waves, created by the sources 1 and 2, the conditions $E=B$ and
${\bf E}\perp {\bf B}$ are fulfilled and therefore the Poynting and
the Umov vectors coincide.

\begin{equation}
{\bf S}_{1}={\bf U}_{1},
\end{equation}

\begin{equation}
{\bf S}_{2}={\bf U}_{2}.
\end{equation}

Let us consider now the resulting electromagnetic field by these two
sources at the same time

\begin{equation}
{\bf E}_{T}={\bf E}_{1}+{\bf E}_{2},
\end{equation}

\begin{equation}
{\bf B}_{T}={\bf B}_{1}+{\bf B}_{2}.
\end{equation}
In this case the Poynting vector and the Umov vector will have the form

\begin{equation}
{\bf S}_{T}=\frac{c}{4\pi }{\bf E}_{T}\times {\bf B}_{T}
={\bf S}_{1}+{\bf S}_{2}+\frac{c}{4\pi }\left({\bf  E}_{1}\times
{\bf B}_{2}+{\bf E}_{2}\times {\bf B}_{1}\right)
\end{equation}
and
\begin{equation}
{\bf U}_{T}=w_{T}c{\bf n},
\end{equation}
where
\begin{equation}
w_{T} =\frac{E_{T}^{2}+B_{T}^{2}}{8\pi }
=w_{1}+w_{2}+\frac{1}{4\pi }\left({\bf E}_{1}\cdot {\bf E}_{2}+
{\bf B}_{1}\cdot {\bf B}_{2}\right)
\end{equation}
and where ${\bf n}$ is the direction of the electromagnetic energy
propagation for the resulting field, that is to say the unitary vector in
the direction of the vector ${\bf S}_{T}$.

The energy flow measured experimentally is the integral over the screen
surface (with normal ${\bf k}$, unitary vector in the positive direction
of the $Z$ axis) of the temporary average of the energy flux density. For each
source  we have
\begin{equation}
\Phi_{1}=\int_{-b}^{b}\int_{-a}^{a}\left\langle {\bf S}_{1}\cdot
{\bf k}\right\rangle _{t}dxdy=\int_{-b}^{b}\int_{-a}^{a}\left\langle
{\bf U}_{1}\cdot {\bf k}\right\rangle _{t}dxdy,
\end{equation}

\begin{equation}
\Phi_{2} =\int_{-b}^{b}\int_{-a}^{a}\left\langle {\bf S}_{2}\cdot
{\bf k}\right\rangle _{t}dxdy=\int_{-b}^{b}\int_{-a}^{a}\left\langle
{\bf U}_{2}\cdot {\bf k}\right\rangle _{t}dxdy,
\end{equation}
and for the resulting field according the Poynting definition

\begin{equation}
\Phi_{P}=\int_{-b}^{b}\int_{-a}^{a}\left\langle {\bf S}_{T}\cdot {\bf k
}\right\rangle _{t}dxdy
\end{equation}
and according Umov definition

\begin{equation}
\Phi_{U}=\int_{-b}^{b}\int_{-a}^{a}\left\langle {\bf U}_{T}\cdot {\bf k
}\right\rangle _{t}dxdy,
\end{equation}
where the designation $\langle\ldots\rangle_t$ is the time average value.

Laborious calculations show that
\begin{equation}
\Phi_{1}=\Phi_{2}=\frac{chE_{0}B_{0}}{8\pi }\int \int \frac{dxdy}{R_{1}^{3}}=
\frac{ chE_{0}B_{0}}{8\pi }\int \int \frac{dxdy}{R_{2}^{3}}\equiv \Phi_{0}
\end{equation}
and that the exact relationships between the resulting flow and the
flows for separate  sources  (${\cal K} =\frac{\Phi}{2\Phi_{0}}$) are
\begin{equation}
{\cal K} _{P}=\frac{\Phi_{P}}{2\Phi_{0}}=1+\frac{\gamma }{2\beta }
\end{equation}
for the Poynting definition, and for the Umov definition
\begin{equation}
{\cal K} _{U}=\frac{\Phi_{U}}{2\Phi_{0}}=\frac{\kappa }{\beta },
\end{equation}
where
\begin{eqnarray}
\gamma &=&\int_{-b}^{b}\int_{-a}^{a}\frac{\cos \left( k\left(
R_{1}-R_{2}\right) \right) }{U_{1}U_{2}R_{1}^{2}R_{2}^{2}}\left[ \frac{
\left( R_{1}+R_{2}\right) R_{2}R_{1}}{r^{2}}\sin ^{2}\theta \right.
\nonumber \\ &&-2\frac{l}{r}\left( R_{2}-R_{1}\right) \sin ^{3}\theta \sin
\varphi \nonumber \\ &&\left. -\left( \frac{l}{r}\right) ^{2}\left(
R_{1}+R_{2}\right) \left( 2\sin ^{2}\theta \sin ^{2}\varphi +\cos
^{2}\theta +2\sin ^{3}\theta \cos ^{2}\varphi \right) \right] dxdy,
\end{eqnarray}

\begin{equation}
\beta =\int_{-b}^{b}\int_{-a}^{a}\frac{dxdy}{R_{1}^{3}},
\end{equation}

\begin{equation}
\kappa =\int_{-b}^{b}\int_{-a}^{a}\int_{0}^{\frac{2\pi }{kc}}\frac{\left(
F+GH\right) \left( J+KL\right) }{\left( A+B+C+2D+2E\right) ^{1/2}}dtdxdy,
\end{equation}

\begin{equation}
A=\frac{1}{R_{1}^{4}}\cos ^{4}\left( kR_{1}-\omega t\right),
\end{equation}

\begin{equation}
B=\frac{1}{R_{2}^{4}}\cos ^{4}\left( kR_{2}-\omega t\right) ,
\end{equation}

\begin{eqnarray}
C &=&\left( \frac{r}{U_{1}U_{2}R_{1}^{2}R_{2}^{2}}\right) ^{2}\cos
^{2}\left( kR_{1}-\omega t\right) \cos ^{2}\left( kR_{2}-\omega t\right)
\nonumber \\ &&\times \left\{ \left[ \left( R_{1}+R_{2}\right)
\frac{R_{2}R_{1}}{r^{2}} \sin ^{2}\theta +\left(
R_{1}U_{2}^{2}-R_{2}U_{1}^{2}\right) \frac{l}{r}\sin \theta \sin \varphi
-\left( R_{2}+R_{1}\right) \left( \frac{l}{r}\right) ^{2}\cos ^{2}\theta
\right] ^{2}+\right. \nonumber  \\ &&\cos ^{2}\theta \left[ \left(
R_{2}-R_{1}\right) \sin ^{2}\theta \sin \varphi +2\frac{l}{r}\left(
R_{2}+R_{1}\right) \sin \theta \cos ^{2}\varphi -\left( \frac{l}{r}\right)
^{2}\left( R_{2}-R_{1}\right) \sin \varphi \right] ^{2}+ \nonumber \\
&&\left.  \cos ^{2}\varphi \left[ \frac{R_{1}R_{2}}{r^{2}}\left(
R_{2}-R_{1}\right) \sin ^{2}\theta -2\frac{l}{r}\left( R_{1}+R_{2}\right)
\cos ^{2}\theta \sin \theta \sin \varphi +\left( \frac{l}{r}\right)
^{2}\left( R_{1}-R_{2}\right) \cos ^{2}\theta \right] ^{2}\right\} ,
\end{eqnarray}

\begin{equation}
D=\frac{r^{2}}{R_{1}^{3}R_{2}^{3}}\left[ 1-\left(
\frac{l}{r}\right) ^{2} \right] \cos ^{2}\left( kR_{1}-\omega t\right)
\cos ^{2}\left( kR_{2}-\omega t\right)
\end{equation}

\begin{eqnarray}
E
&=&\frac{r^{2}}{U_{1}U_{2}R_{1}^{2}R_{2}^{2}}\cos \left( kR_{1}-\omega
t\right) \cos \left( kR_{2}-\omega t\right) \times \left[
\frac{1}{R_{1}^{2}} \cos ^{2}\left( kR_{1}-\omega t\right)
+\frac{1}{R_{2}^{2}}\cos ^{2}\left( kR_{2}-\omega t\right) \right]
\nonumber \\ &&\times \left[ \left( 1+\frac{R_{1}R_{2}}{r^{2}}\right) \sin
^{2}\theta -\left( \frac{l}{r}\right) ^{2}\left(
1+\frac{R_{1}R_{2}}{r^{2}}-3\sin ^{2}\theta +4\sin ^{2}\theta \sin
^{2}\varphi \right) +\left( \frac{l}{r} \right) ^{4}\right]
\end{eqnarray}

\begin{equation}
F=\frac{1}{R_{1}^{3}}\cos ^{2}\left( kR_{1}-\omega t\right) +\frac{1}{
R_{2}^{3}}\cos ^{2}\left( kR_{2}-\omega t\right) ,
\end{equation}

\begin{equation}
G=\frac{1}{U_{1}U_{2}R_{1}^{2}R_{2}^{2}}\cos \left( kR_{1}-\omega t\right)
\cos \left( kR_{2}-\omega t\right) ,
\end{equation}

\begin{eqnarray}
H &=&\left( R_{1}+R_{2}\right) \frac{R_{2}R_{1}}{r^{2}}\sin ^{2}\theta
-2\left( R_{2}-R_{1}\right) \frac{l}{r}\sin ^{3}\theta \sin \varphi
\nonumber \\ &&-\left( \frac{l}{r}\right) ^{2}\left( R_{1}+R_{2}\right)
\left[ 2\sin ^{2}\theta \sin ^{2}\varphi +\cos ^{2}\theta +2\sin
^{3}\theta \cos ^{2}\varphi \right],
\end{eqnarray}

\begin{equation}
J=\frac{1}{R_{1}^{2}}\cos ^{2}\left( kR_{1}-\omega t\right) +\frac{1}{
R_{2}^{2}}\cos ^{2}\left( kR_{2}-\omega t\right),
\end{equation}

\begin{equation}
K=\frac{r^{2}}{U_{1}U_{2}R_{1}^{2}R_{2}^{2}}\cos \left( kR_{1}-\omega
t\right) \cos \left( kR_{2}-\omega t\right),
\end{equation}

\begin{equation}
L=\left( 1+\frac{R_{1}R_{2}}{r^{2}}\right)  \sin ^{2}\theta -\left(
\frac{l}{r }\right) ^{2}\left( 1+\frac{R_{1}R_{2}}{r^{2}}-3\sin ^{2}\theta
+4\sin ^{2}\theta \sin ^{2}\varphi \right) +\left( \frac{l}{r}\right)
^{4},
\end{equation}

\begin{equation}
U_{1}=\left( \sin ^{2}\theta +2\frac{l}{r}\sin \theta \sin
\varphi +\frac{ l^{2}}{r^{2}}\right) ^{\frac{1}{2}},
\end{equation}

\begin{equation}
U_{2}=\left(
\sin ^{2}\theta -2\frac{l}{r}\sin \theta \sin \varphi +\frac{
l^{2}}{r^{2}}\right) ^{\frac{1}{2}},
\end{equation}

\begin{equation}
R_{1}^{2}=r^{2}+l^{2}+2lr\sin \theta \sin \varphi,
\end{equation}

\begin{equation}
R_{2}^{2}=r^{2}+l^{2}-2lr\sin \theta \sin \varphi.
\end{equation}

The ${\cal K} _{P}$ and ${\cal K} _{U}$ values as a function of the angle $\alpha$
between the rays, were obtained numerically for the  values
$\sqrt{l^2+h^2}=0.3 \,{\tt m}$, $\lambda =632.8\, {\tt nm}$, $a=3.5\, {\tt
mm}$  and $b=2.5\, {\tt mm}$  used in the experiment.  The graph of these
dependencies are presented in the Fig. 2.

\begin{figure}
\caption{Theoretical curves of the behavior of the coeficient ${\cal K}$ versus angle
according the Poynting and Umov vectors.}
\end{figure}

It is not difficult to show that the previous results for the relationship
between the resulting flow and the flows for separate sources
(${\cal K} =\frac{\Phi}{2\Phi_{0}}$) are conserved if the spherical waves 1 and 2
are modeled by means of the equations

\begin{equation}
{\bf E}_{1}
=\frac{E_{0}}{R_{1}}\cos \left( {\bf k}_{1}\cdot
{\bf R}_{1}-\omega t\right) {\bf e}_{\varphi 1},
\end{equation}

\begin{equation}
{\bf B}_{1} =-\frac{B_{0}}{R_{1}}\cos \left( {\bf k}_{1}\cdot
{\bf R}_{1}-\omega t\right) {\bf e}_{\theta 1},
\end{equation}

\begin{equation}
{\bf E}_{2} =\frac{E_{0}}{R_{2}}\cos \left( {\bf k}_{2}\cdot
{\bf R}_{2}-\omega t\right) {\bf e}_{\varphi 2},
\end{equation}

\begin{equation}
{\bf B}_{2} =-\frac{B_{0}}{R_{2}}\cos \left( {\bf k}_{2}\cdot
{\bf R}_{2}-\omega t\right) {\bf e}_{\theta 2}.
\end{equation}

\section{Description of the experiment}

The diagram of the experimental arrangement is shown in the Fig.3

\begin{figure}
\caption{The experimental arrangement diagram. I1 and I2 are optical
obturators. ADC is the analog digital converter.}
\end{figure}

As coherent monochrome light source  a {\tt He-Ne}  laser model
08181.93 from the company ``PHYWE" with  parameters: wavelength  $l = 632.8
{\tt nm}$, beam power $P = 1 {\tt mW}$, polarization $500:1$,  beam
diameter $d = 0.5 {\tt mm}$, is used.  The laser beam goes to the
beamsplitter $B1$, where it is unfolded in two rays of the same intensity
approximately. The reflected ray goes to the reference photodiode PIN1,
model 1PP75 from the company ``TESLA", that works in short circuit regime.
The photodiode-generated current is amplified by an amplifier DC and by
 means of an analogical-digital converter ADC  is received on the computer
PC.  This signal serves as a reference signal and gives us information
about the intensity changes of the laser beam.

The part of the laser beam, which goes through the   beamsplitter B1, goes
to the second beamsplitter B2, where on the other  hand it is unfolded in
two rays. The second-beamsplitter-reflected ray goes to the optic
obturator I2, then reflected by the mirror M2 and  falls on the lens
L2.  The lens L2 is a biconvex lens and possesses focal  distance of
$18 {\tt mm}$. This lens transforms the cylindrical ray in a divergent
beam.  After going through the lens the beam goes to the measuring
photodiode PIN2 (type 1PP75 from "TESLA") and it falls on its active
surface under  an angle $\alpha$.  The ray that goes through the
beamsplitter B2, goes consecutively through the optic obturator I1 and the
adjustable compensator C, then is reflected  by the mirror M1 and falls on
the lens L1. The lens L1 is of the same type  as the lens L2 and has the
same function. After going through the lens L1 the  divergent beam,
produced by the lens L1, falls on the measuring photodiode  under the same
angle.  The angle between both beams falling on the PIN2  is equal to
$2\alpha$ (Fig.3).  The photodiode PIN2 also works in short circuit
regime.  Its signal is amplified by another amplifier DC and goes through
an analogical-digital converter toward the computer. Its signal is
proportional to the luminous flow that falls on its  active surface.

The mirrors M1 and M2 are movable. The distances between each lenses
and the measuring photodiode PIN2 are the same and they measure $30
{\tt cm}$.  These distances stay fixed during the experiment. The angle
$\alpha$ is changed only from $14^o$ until $86^o$ and its value is
measured within the accuracy of  $0.5^o$. When carrying out the
experiment the result of each measurement is corrected with the reading of
the photodiode PIN1.  By this way the error produced by the laser
instability is avoided.

All the measurements are carried out  in a dark room. To avoid the
influence of the laser instability on  the experimental results,
 a  normalization of the readings of the photodiode PIN2
is executed  with  the help of photodiode PIN1 for  all the measurements.
For each value of the angle $\alpha$ the experiment is executed in three
stages:

{\bf Stage 1}: Both optical obturators I1 and I2 are closed up,
and by means of the photodiode PIN2 the ground is measured. As  the
ground value was always  below 0.5\% of all other measurements, any
correction is not applied to the experimental results.

{\bf Stage 2}:  This stage has as a goal to equal and to measure the light
energy flows that  go through both optical branches in the experimental
arrangement:  branch $1$  (optical obturator I1, adjustable compensator
C, mirror M1, lens $L1$) and branch  $2$ (optic obturator I2, mirror
M2, lens L2). The obturator I1 closes up and  the obturator I2 opens
up.  The photodiode-PIN2-generated current is  measured. Then the
obturator I1 opens up and the obturator I2 closes up.  By means of the
adjustable compensator C the present current in the photodiode  PIN2 is
adjusted similarly to its previous current with an error limited  to  1\%.
By this way the readings of the photodiode PIN2 that correspond to the
optical flow $\Phi_1$ that passes through the branch $1$,  and also
correspond to  the optical flow $\Phi_2$, that passes through the branch $2$,
are already known.

{\bf Stage 3}: Both obturators  I1 and I2 open up and the current of
the photodiode PIN2 that represents the total flow $\Phi$ is
measured. Then the computer PC calculates the coefficient

\begin{equation}
{\cal K}=\frac{\Phi}{\Phi_1+\Phi_2},
\end{equation}
and it memorizes these values as a function of the angle $\alpha$.

The experiment was carried out for the two light flows configurations
shown in the figs. 4-5,

\begin{figure}
\caption{Equivalent scheme of the experimental arrangement when
the electric field  vectors from the wave in the branch  1 are parallel to
the electric field vectors from the wave in the branch 2,  $2\alpha$ is
the angle between straights connecting each of two sources with the centre
of the sensor} \end{figure}

\begin{figure}
\caption{Equivalent scheme of the experimental arrangement when
the magnetic field  vectors from the wave in the branch  1 are parallel to
the magnetic field vectors from the wave in the branch 2.} \end{figure}

\noindent
which correspond to the Eqs. (10)-(13) and (50)-(53). The experimental
results are presented in the graphs (fig. 6 and fig. 7).

\clearpage

\begin{figure}
\caption{The theoretical curves of the behavior of the coeficient
 ${\cal K}$
versus angle according to the Poynting vector and according to the Umov
vector and the experimental results for the case of the waves with
parallel magnetic field vectors.}
\end{figure}

\begin{figure}
\caption{The theoretical curves of the behavior of the coeficient
${\cal K}$
versus angle according to the Poynting vector and according to the Umov
vector and the experimental results for the case of the waves with
parallel electric field vectors. }
\end{figure}

\section{Conclusions}

Our work  starts from the position that the Umov vector defines,  in a
general way, the  energy flow for {\it any} type of energy and it is a
consequence of the  energy conservation law. Its expression for the
particular case of  electromagnetic waves is ${\bf U}=wc{\bf n}$. On the
other hand the  Poynting vector defines the flow of the electromagnetic
energy as a  consequence of the energy conservation law in the Maxwell's
theory.

However,  although  both these vectors represent the same physical quantity
and therefore  they {\it must} coincide, we demonstrated  that their
equality is  limited to the case when the fields, generating the
electromagnetic  energy flow, fulfill the conditions $E=B$ (in CGS) and
${\bf E}\bot{\bf B}$.

Obviously these  conditions are not general and therefore one can find
situations when  the electromagnetic field does not fulfill one or both
these conditions.  In order to determine which of the definitions of the
energy density flux  vector (Umov's vector or Poynting's vector), in the
case when these conditions  are not fulfilled, gives correct prediction
about the energy flow,  in the  present paper we experimentally measured
the flow of the resultant  electromagnetic energy  of two bunches
of electromagnetic waves when  aforementioned conditions are not
fulfilled.  This experiment shows that  the Umov vector does not describe
appropriately the electromagnetic  energy flow, while the Poynting vector
{\it does}.

Therefore, the experiment apparently  shows that the energy flux density
definition through the Umov vector  {\it in general} is not applicable to
electromagnetic phenomena. However  {\it such} conclusion  {\it must be}
necessarily incorrect. Indeed, the expression for the
Umov vector is obtained starting from  the universally  accepted
conservation law of any type of energy.  Consequently the Umov and Poynting vectors should always coincide. For this reason
there is an apparent contradiction in the  fact that the flow
theoretically predicted  on the basis of  Umov vector   coincides neither
with  the experimentally measured flow  nor  with the  corresponding
flow calculated  by means of the Poynting vector.

The explanation for this apparent  contradiction can reside in the
following:

Because the Umov vector is a consequence of the energy conservation law,
 the Umov vector functional dependence $({\bf U}=wc{\bf n})$ should be
correct and therefore it is necessary to examine the elements used in its
construction, namely, $w$ (energy density), $c$ and ${\bf n}$.

The propagation velocity of the electromagnetic waves (and consequently,
of the electromagnetic energy) in vacuum is unique and is equal to $c$,
and therefore there is no problem.

However, on the one hand, if the standard definition of the Poynting
vector is correct it defines the correct direction of the
electromagnetic energy flux and, in turn, obviously, the direction of
the Umov vector, then the standard expression of the electromagnetic
density utilized  to build the Umov vector  {\it cannot be} correct. On
the other hand if the standard definition of the energy density is correct
then the unit vector ${\bf n}$ (direction of the Poynting vector) used for
constructing the Umov vector must be incorrect and therefore the direction
of the energy  propagation is not correct. This would  mean that the
standard expression  for the Poynting vector does not define correctly
the electromagnetic energy density flux. Note, however, that our
calculations and experiment in which we used the standard expression of
the Poynting vector do not contradict the last claim: the point is that
the Poynting vector is defined with the accuracy of the curl of an arbitrary vector
and for this reason, in principle, it is possible that the same result
for the integral (27) will be obtained from another expression for the
electromagnetic flux density. If this is the case then the direction of
the energy flow does not have the direction of ${\bf E}\times{\bf B}$.

So for the case examined by us there is an incompatibility between the
generally accepted definition  of the electromagnetic energy density and
the conventional definition of the energy flux density expressed by the
Poynting vector. This particular case allows us to affirm that, in
general, these standard definitions {\it are incompatible}.

\acknowledgments
The authors would like to express their gratitude to  the Rector of the
University of Zacatecas   Rogelio Cardenas Hernandez  for constant
support. We would also like to thank Ernesto Mendivil  Barreras for
revising the manuscript.

\end{document}